\documentclass[a4,twocolumn,amsmath,amssymb,aps,prl,superscriptaddress]{revtex4}
\usepackage{graphicx}
\usepackage{amsfonts}
\usepackage{amsmath}
\usepackage{amssymb}
\begin{document}

\title{Transform-Limited X-Ray Pulse Generation from a \\High Brightness Self-Amplified Spontaneous Emission Free-Electron Laser}

\author{B. W. J. M$^{\mathrm{c}}$Neil}
\email{b.w.j.mcneil@strath.ac.uk} 
\affiliation{University of
Strathclyde (SUPA), Glasgow G4 0NG, UK}

\author{N. R. Thompson}
\email{neil.thompson@stfc.ac.uk}
\affiliation{University of Strathclyde
(SUPA), Glasgow G4 0NG, UK} \affiliation{ASTeC, Daresbury
Laboratory, Warrington WA4 4AD, UK.}
\author{D. J. Dunning}
\email{david.dunning@stfc.ac.uk} 
\affiliation{University of Strathclyde
(SUPA), Glasgow G4 0NG, UK} \affiliation{ASTeC, Daresbury
Laboratory, Warrington WA4 4AD, UK.}

\date{\today}

\begin{abstract}
A method to achieve High-Brightness Self-Amplified Spontaneous Emission (HB-SASE) in the Free Electron Laser (FEL) is described. The method uses repeated non-equal electron beam delays to de-localise the collective FEL interaction and break the radiation coherence length dependence on the FEL cooperation length. The method requires no external seeding or photon optics and so is applicable at any wavelength or repetition rate. It is demonstrated using linear theory and numerical simulations that the radiation coherence length can be increased by approximately two orders of magnitude over SASE with a corresponding increase in spectral brightness. Examples are shown of HB-SASE generating transform-limited FEL pulses in the soft X-ray and near transform-limited pulses in the hard X-ray. Such pulses may greatly benefit existing applications and may also open up new areas of scientific research.
\end{abstract}

\maketitle

In the X-Ray region of the spectrum  Self-Amplified Spontaneous Emission (SASE) free-electron lasers (FELs) are currently opening up new frontiers across science \cite{FLASH,LCLS,SACLA1,SACLA2,FERMI@elettra,XFELSREVIEW}.
Although SASE FELs have brightness up to $10^8$ times greater than laboratory sources their full potential is limited by a relatively poor temporal coherence.
In this Letter, a High Brightness SASE (HB-SASE) FEL is described which may significantly improve the temporal coherence towards the transform limit, enhancing the spectral brightness and potentially enabling X-ray FELs to enhance their existing scientific capability and to access new experimental regimes.

Applications that may benefit include Resonant Inelastic X-ray Scattering (RIXS) \cite{RIXS} which requires substantial incident photon flux to collect sufficient spectra with a high enough resolution in energy and momentum in a reasonable time. RIXS has evolved greatly over the last decades, due to increases in photon flux from synchrotron sources then X-ray FELs, as well as advances in instrumentation.
Each increase in resolution has revealed new details in material excitation spectra. Further improvement may extend study to e.g. single magnons in small exchange systems and the full dispersion curves of superconducting gaps. New applications made possible by fourier-transform limited X-ray pulses may include time-resolved X-ray spectroscopy of chemical dynamics, or quantitative studies of molecular and cluster fragmentation, where high repetition rate pulses of controlled temporal profile are essential for systematic studies of nonlinear phenomena \cite{NLS}.

In the SASE FEL \cite{SASE2, SASE1}, a relativistic bunch of electrons with a Lorentz factor of $\gamma\gg1$ enters an undulator comprising an array of transverse, alternating polarity dipole magnets of period $\lambda_u$.
The undulator magnetic fields cause the electrons to oscillate transversely and emit initially incoherent radiation.
A cooperative (or collective) instability in the coupled electron-radiation system may cause an exponential gain in the radiation field and an electron microbuncing at the resonant radiation wavelength $\lambda_r=\lambda_u(1+\bar{a}_u^2)/2\gamma^2$, where $\bar{a}_u$ is the rms undulator parameter, proportional to the undulator period and magnetic field.
The interaction is a positive feedback process  --- the electron-radiation coupling drives the radiation phase giving a greater electron microbunching at the resonant wavelength and greater coherent radiation emission~\cite{XFELSREVIEW}. In the linear regime the power growth is given by $P(z) \sim P_0/9 \exp(\sqrt{3}z/l_g)$ where the nominal  gain length $l_g\equiv\lambda_u/4\pi\rho$ with $\rho$ the dimensionless FEL parameter~\cite{SASE2}.
The process  leaves the linear regime and saturates when the electrons become strongly bunched at the radiation wavelength and then begin to de-bunch.

This type of collective process also describes several other particle-radiation interactions, such as the collective atomic recoil laser~\cite{CARL}, collective Rayleigh scattering from linear dielectric particles~\cite{RAYLEIGH} or collective scattering from the electron-hole plasma in semiconductors~\cite{HOLE}.
A significant difference between the FEL  and these latter interactions is the relativistic speed of the electron bunch as it propagates through the FEL.
In each undulator period $\lambda_u$ a wavefront at resonant radiation wavelength $\lambda_r$ propagates through the electron beam a distance $\lambda_r$ --- this is referred to as `slippage'.
On propagating through one gain length of the undulator, a wavefront propagates through the electron beam a distance $l_c\equiv\lambda_r/4\pi\rho\ll l_g$.
This `cooperation length' $l_c$ therefore defines the scale at which collective effects evolve throughout the electron beam, and so how the temporal coherence of the radiation field evolves from the initially spontaneous noise.
For a sufficiently long electron beam, different regions along the beam develop from the localised noise source autonomously and are therefore uncorrelated in phase. In this sense, the SASE process can be considered as a `localised' collective process.
At saturation the SASE output then comprises a series of phase-uncorrelated `spikes' which can be shown to be separated by $\lesssim 2\pi l_c$~\cite{SASELC}. Typically, in the X-ray the electron bunch length $l_b \gg 2\pi l_c$, so the output comprises many such spikes and the pulse is far from fourier-transform limited and consequently of reduced spectral brightness from that potentially available.

Several methods may be used to improve the temporal coherence of the SASE FEL output.
These can be divided into two general classes.
In the first class, an externally injected source of good temporal coherence `seeds' the FEL interaction so that noise effects are reduced. This seed field may be either at the resonant radiation wavelength, where available, or at a subharmonic which is then up-converted within the FEL. These methods, which include High Gain Harmonic Generation (HGHG)~\cite{HGHG1,HGHG2,HGHG3,HGHG4} and Echo-Enabled Harmonic Generation (EEHG)~\cite{EEHG1,EEHG2}, rely on a synchronised external seed at the appropriate wavelength, pulse energy and repetition rate.
In the second class, the coherence is created by optical manipulation of the FEL radiation itself, for example by spectrally filtering the SASE emission at an early stage for subsequent re-amplification to saturation in a self-seeding method~\cite{SELFSEEDING1,SELFSEEDING2,SELFSEEDING3,SELFSEEDING4}, or via the use of an optical cavity~\cite{XFELO1,XFELO2,RAFEL1,RAFEL2,RAFEL3,RAFEL4,RAFEL5,RAFEL6}. Methods in this class rely on potentially complex material-dependent optical systems which limit the ease and range of wavelength tuning. If an optical cavity is used, the electron source repetition rate should also be in the MHz regime to enable a practical cavity length.

The HB-SASE method described in this Letter requires no external seeding or photon optics and is thus applicable at any wavelength and repetition rate. It works by using a series of magnetic chicanes inserted between each section of a long sectional undulator.
The chicanes periodically delay the electron bunch with respect to the co-propagating radiation field.
The delays have two main effects.
Firstly, a radiation wavefront propagates at a rate through the electron beam which can be significantly greater than in SASE -- the slippage rate is increased.
Secondly, if the undulator sections are shorter than the gain length $l_g$, and the chicanes introduce delays which are greater than the cooperation length $l_c$, then the localised nature of the collective interaction may be broken, so inhibiting the formation of the phase-uncorrelated radiation spikes associated with SASE.
The electron-radiation interaction distance over which temporal coherence can be developed therefore becomes much greater than the $\lesssim 2\pi l_c$ of SASE, and potentially towards the total length that a resonant wavefront propagates through the electron beam to saturation. This process is aided by the relatively fast rate of change of the radiation phase that can occur in the linear regime~\cite{XFELSREVIEW} and which allows the temporal coherence to propagate via a radiation phase which evolves with much greater uniformity than can occur in SASE.

An initial study over a limited parameter range~\cite{ILONGC} demonstrated the basic principle of improving the temporal coherence but it is now shown in this Letter that by de-localising the collective interaction as described above the coherence length of the radiation may in fact be extended by orders of magnitude with a corresponding enhancement in the spectral brightness. Hence the name High-Brightness SASE (HB-SASE).

The potential of HB-SASE is demonstrated using a FEL numerical simulation code that solves the universally scaled one-dimensional FEL equations~\cite{FELO}
\begin{eqnarray}
\frac{d \theta_j}{d\bar{z}}&=&p_j \label{tdot} \\
\frac{d p_j}{d\bar{z}}&=&-\left(A\left(\bar{z},\bar{z}_1\right)\exp \left(i\theta_j\right)+c.c.\right)\label{pdot} \\
\left(\frac{\partial}{\partial \bar{z}}\right.&+&\left.\frac{\partial}{\partial \bar{z}_1}\right)A\left(\bar{z},\bar{z}_1\right)=
\chi\left(\bar{z}_1\right)b\left(\bar{z},\bar{z}_1\right)\label{adot}.
\end{eqnarray}
In this scaling the independent parameters are the combined interaction distance through the undulator which is scaled by the gain length, $\bar{z}={z}/{l_g}$, and the distance in the electron beam rest frame which is scaled by the cooperation length, $\bar{z}_1={(z-c\bar{\beta}_z t)}/{l_c}$. The dependent electron-radiation parameters are defined as follows:
$\theta_j=\left(k+k_u\right)z-\omega t_j$ is the ponderomotive phase of the $j$th electron;
$p_j=\left(\gamma_j-\gamma_r\right)/\rho\gamma_r$ is its scaled energy;
$b\left(\bar{z},\bar{z}_1\right)\equiv \left.\left<e^{-i\theta\left(\bar{z}\right)}\right>\right|_{\bar{z}_1} $ is the electron bunching factor for which
$0\leq |b|<1$ and is the average over the electrons contained within the ponderomotive well centred at $\bar{z}_1$ at distance through the interaction region $\bar{z}$;  $\chi\left(\bar{z}_1\right)=I\left(\bar{z}=0,\bar{z}_1\right)/I_{pk}$ is the current weight factor where $
I\left(\bar{z}=0,\bar{z}_1\right)$ describes the electron pulse current distribution of peak value $I_{pk}$ at the entrance to the FEL interaction region; $A\left(\bar{z},\bar{z}_1\right)$ is the scaled complex radiation field envelope of magnitude approximated by
$\left|A\right|^2\sim P_{rad}/\rho P_{beam}$ with $P_{rad}$ and $ P_{beam}$ the radiation and peak electron beam powers respectively.
The scaled frequency is $\bar{\omega}=(\omega-\omega_{r})/2\rho\omega_{r}$.

The effect of a chicane is to delay the electrons by a longitudinal shift of $\bar{\delta}$, so that $\bar{z}_1\rightarrow \bar{z}_1-\bar{\delta}$.
Radiation wavefront propagation in $\bar{z}_1$ within each undulator section is a constant $\bar{l}$, so the total relative displacement for the $n$th undulator-chicane module is $\bar{s}_n=\bar{l}+\bar{\delta}_n$.
For equal delays ($\bar{\delta}_n=\bar{\delta}$, a constant) the system is identical to that of the Mode-Coupled FEL \cite{MODELOCKEDFEL}, where the  frequency spectrum displays discrete sideband modes at spacing $\Delta \bar{\omega}=2\pi/\bar{s}$ and the field in the temporal domain is strongly modulated with period $\bar{s}=\bar{l}+\bar{\delta}$.
In HB-SASE however the chicane delays of the $n$th undulator-chicane module $\bar{\delta}_n$, are not equal.
With an appropriately chosen sequence of unequal delays, the set of sideband mode frequencies supported by each undulator-chicane module can be made  unique.
Only those frequencies immediately about the resonant frequency $\bar{\omega}=0$ will then receive amplification---the process may thus be considered as a sequential filtering of unwanted frequencies, or a self-seeding of the growing radiation field.

The FEL equations (\ref{tdot},\ref{pdot},\ref{adot}) can be linearised using collective variables~\cite{LIN}, and solved in the frequency domain via a Fourier-Laplace transformation. The solution for the interaction through to the end of the $n$th undulator-chicane module is obtained by sequentially applying the linear solution using the output of module $(n-1)$ as the initial conditions for the $n$th module to  give:
\begin{equation}\label{linear_spectrum}
\tilde{x}_j^{(n)}=-e^{{i \bar{\omega}\bar{\delta}_j^{(n-1)}}}\sum_{k=1}^3 \tilde{x}_k^{(n-1)}\sum_{p=1}^3\frac{a_{jk}(\lambda_p)e^{i\lambda_p\bar{l}}}
{\prod_{q\neq p}(\lambda_p-\lambda_q)}
\end{equation}
where
\begin{equation}
\tilde{x}^{(n)}=
\begin{bmatrix}
\tilde{b}^{(n)}\\
\tilde{P}^{(n)}\\
\tilde{A}^{(n)}
\end{bmatrix},
\quad
a_{jk} = \begin{bmatrix}
-\lambda(\lambda+\bar{\omega}) & (\lambda+\bar{\omega}) & i \\
1 & -\lambda(\lambda+\bar{\omega}) & -i\lambda \\
i\lambda & -i & -\lambda^2
\end{bmatrix},
\end{equation}
where $\lambda_{p,q}$ are roots of the characteristic equation $\lambda^3+\bar{\omega}\lambda^2+1=0$ and a $\bar{\omega}$ dependence has been assumed on the variables.
Here, the electron variables at the end of the $n$th undulator section are shifted back in $\bar{z}_1$ by the chicane using the Fourier shift theorem by multiplying them by $\exp(i\bar{\omega}\bar{\delta}_j^n)$, so that $\bar{\delta}_j^n=\bar{\delta}^n$ for $j=1,2$ and $\bar{\delta}_j^n=0$ for the field variable with $j=3$.

The total gain bandwidth envelope of (\ref{linear_spectrum}) is the single undulator section spectrum, as for the mode coupled case of~\cite{MODELOCKEDFEL}, i.e. it is a sinc function with first zero at $\bar{\omega}=\pm 2\pi/\bar{l}$.
The linear solution can now be used to optimise a sequence of chicane delays.
For example, the sequence used in the results presented in this Letter is given by $\bar{s}_n=\mathbb P_n \bar{s}_1/2$ where $\bar{s}_1=\bar{l}+\bar{\delta}_1$ and $\mathbb P_n$ are a series of primes with $\mathbb P_1=2$.
 It is found that, for this sequence, in order for no common modes to exist within the gain bandwidth $\bar{s}_1\leq 2\bar{l}$ must be satisfied, and similarly for no common modes within the FWHM bandwidth $\Delta \bar{\omega}_{\mathrm{FWHM}} \simeq 5.4/\bar{l}$~\cite{MODELOCKEDFEL}, it is necessary for $\bar{s}_1<4.65\bar{l}$.
With some  empirical adjustment it may be possible to optimise further this particular delay sequence and circumvent these limits.
Other sequences of delays (for example random variation about a set mean or quasiperiodic \cite{QUASIPERIODIC}) have also been shown in simulations to be effective, and can be optimised in a similar way using~(\ref{linear_spectrum}).

An estimate from~(\ref{linear_spectrum}), valid for $\bar{l}=0.5$ and any sequence of chicane delays, shows that the FWHM bandwidth around the resonant frequency $\bar{\omega}=0$ is given by  $\Delta \bar{\omega}\simeq 4\pi/\bar{S}$ where $\bar{S}\equiv\sum_n \bar{s}_n$ and is thus inversely proportional to the total slippage.
Using the time bandwidth relation  $\Delta \nu \Delta t \gtrsim 1$ (in scaled units, $\Delta \bar{\omega} \Delta \bar{z}_1 \gtrsim 2\pi$) the \emph{maximum} coherence length predicted by the linear theory is $\bar{l}_{\mathrm{coh}}\simeq \bar{S}/2$.

\begin{figure}[htb]
\includegraphics[width=90mm]{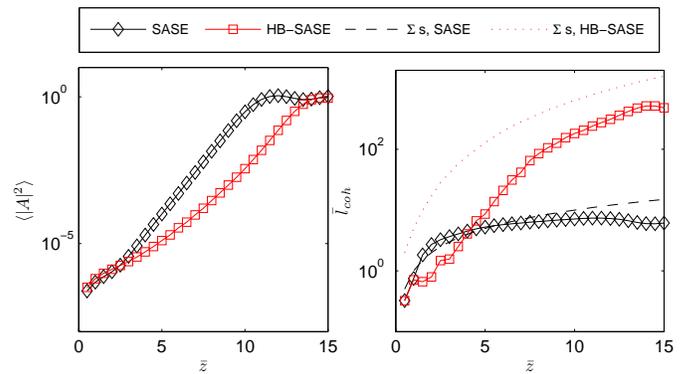}
\caption{Evolution with propagation distance $\bar{z}$ of the scaled optical power $\langle |A|^2\rangle$, and radiation coherence length $\bar{l}_{\mathrm{coh}}$. Also shown is the accumulated total slippage $\bar{S}(\bar{z})$.}\label{squaredata}
\end{figure}

HB-SASE has been simulated using the above model for a range of parameters.
A typical example is now shown which demonstrates the large increase in radiation coherence length of HB-SASE over normal SASE.
The electron beam is long, of scaled length $\bar{l}_e=4000$  and of constant current, i.e. $\chi(\bar{z}_1)=1$. The electrons are cold (no energy spread), the scaled undulator section lengths are $\bar{l}=0.5$ and a sequence of chicane delays used as above with $\bar{s}_1=4\bar{l}$. Effects arising from emission at the rear of the electron pulse have been omitted.
Figure \ref{squaredata} plots as a function of $\bar{z}$ the scaled optical power averaged over the pulse $\langle|A|^2 \rangle$, and the scaled radiation coherence length $\bar{l}_{\mathrm{coh}}(\bar{z}_1)=\int^{\infty}_{-\infty} |g(\bar{\tau}_1)|^2 d\bar{\tau}_1$, with $g(\bar{\tau}_1)= \langle A^*(\bar{z}_1)A(\bar{z}_1+\bar{\tau}_1)\rangle/ \langle A^*(\bar{z}_1)A(\bar{z}_1) \rangle$~\cite{LCOHDEFN}.
Also shown is an equivalent SASE simulation without chicane delays (i.e. $\bar{\delta}_n=0 \;\forall\;  n$). For SASE, the saturation length is $\bar{L}_{\mathrm{sat}}(\bar{z})=12.0$, the coherence length $\bar{l}_{\mathrm{coh}}(\bar{z}_1)=7.3$ (close to the SASE spike spacing of $2\pi$ in units of $\bar{z}_1$) and the rms bandwidth $\sigma_{\bar{\omega}}=0.5$ in agreement with SASE behaviour~\cite{SASELC}.
For HB-SASE the saturation length is slightly increased to $\bar{L}_{\mathrm{sat}}(\bar{z})=14.5$ with mean saturated power similar to SASE.
However, the coherence length of HB-SASE has increased by almost two orders of magnitude to $\bar{l}_{\mathrm{coh}}(\bar{z}_1)=517$ with a bandwidth correspondingly reduced to $\sigma_{\bar{\omega}}=0.0045$, to give an increase in spectral brightness of nearly two orders of magnitude over SASE. It is seen that the SASE coherence length $\bar{l}_{\mathrm{coh}}$, evolves little after the first three gain lengths, reaching half its saturation value by $\bar{z}=3$ and thereafter growing more slowly than the accumulated slippage $\bar{S}$, tending asymptotically to $\bar{l}_{\mathrm{coh}}\simeq 2\pi$. This is consistent with the localised growth of the radiation spikes uncorrelated in phase. Conversely, for HB-SASE, $\bar{l}_{\mathrm{coh}}$ evolves little for the first three gain lengths, then grows exponentially with a growth rate greater than $\bar{S}$. This appears consistent with a more de-localised interaction as discussed above, where initially the coherence takes longer to develop, but then becomes more established as the radiation phase changes rapidly towards a common value over a range approaching the accumulated slippage $\bar{S}$. When the radiation phase is plotted it is seen to vary smoothly and slowly.

\begin{figure}[hbt]
\includegraphics[width=38mm]{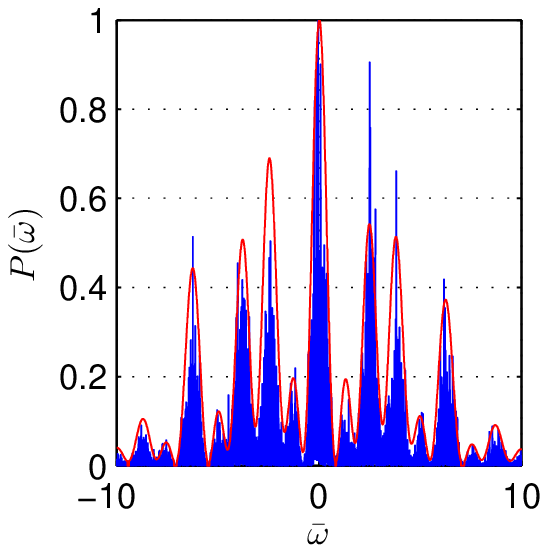}
\includegraphics[width=38mm]{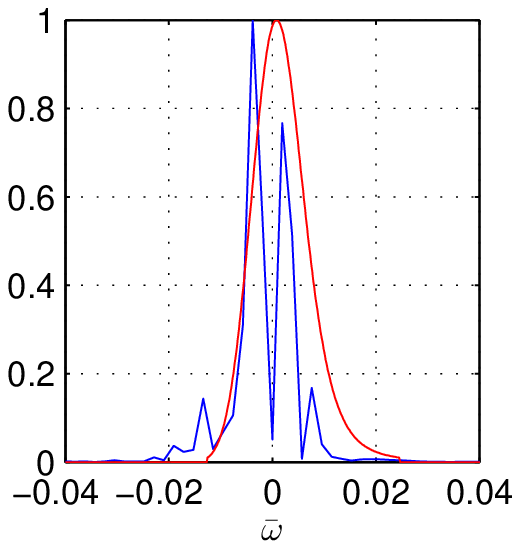}
\caption{Comparison of the radiation spectra from the simulation results (blue) and by the linear theory (\ref{linear_spectrum}) (red), at $\bar{z}=1.5$ (left) and $\bar{z}=12.5$ (right). Plots are scaled with respect to their peak values.}\label{lineartheory}
\end{figure}

Agreement between the radiation spectra from the simulation results and that predicted by the linear theory of~(\ref{linear_spectrum}) is shown in Figure~\ref{lineartheory}, near the beginning of the interaction at $\bar{z}=1.5$ (top) and at the end of the linear regime at $\bar{z}=12.5$ (bottom). At $\bar{z}=12.5$ the FWHM bandwidth predicted by the linear theory is
$\Delta \bar{\omega}=0.011$. The predicted coherence length, using $\bar{l}_{\mathrm{coh}} \simeq 2\pi/\Delta \bar{\omega}$ is $\bar{l}_{\mathrm{coh}}=571$ in good agreement with $\bar{l}_{\mathrm{coh}}=446$ for the simulation optical field.

\begin{figure}[h]
\includegraphics[width=42mm]{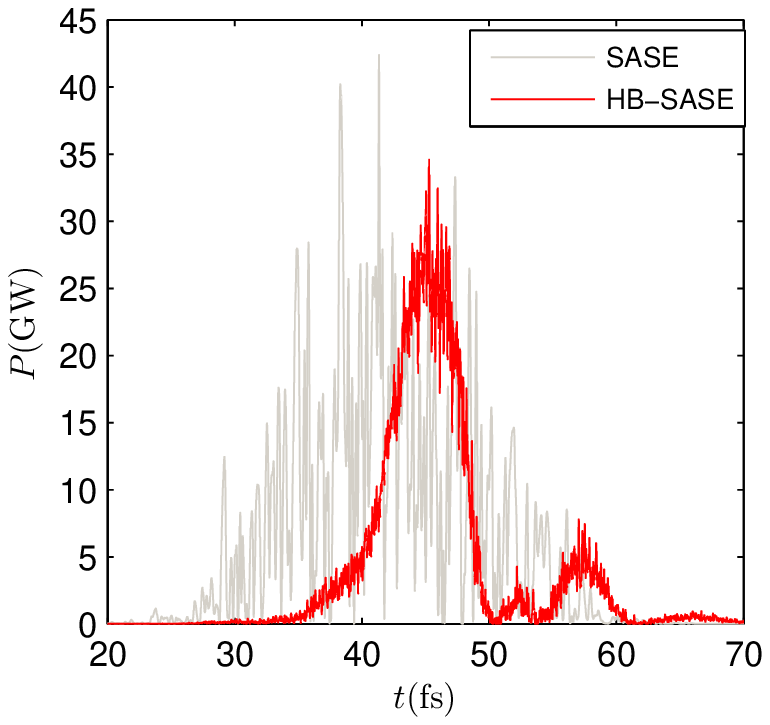}
\includegraphics[width=42mm]{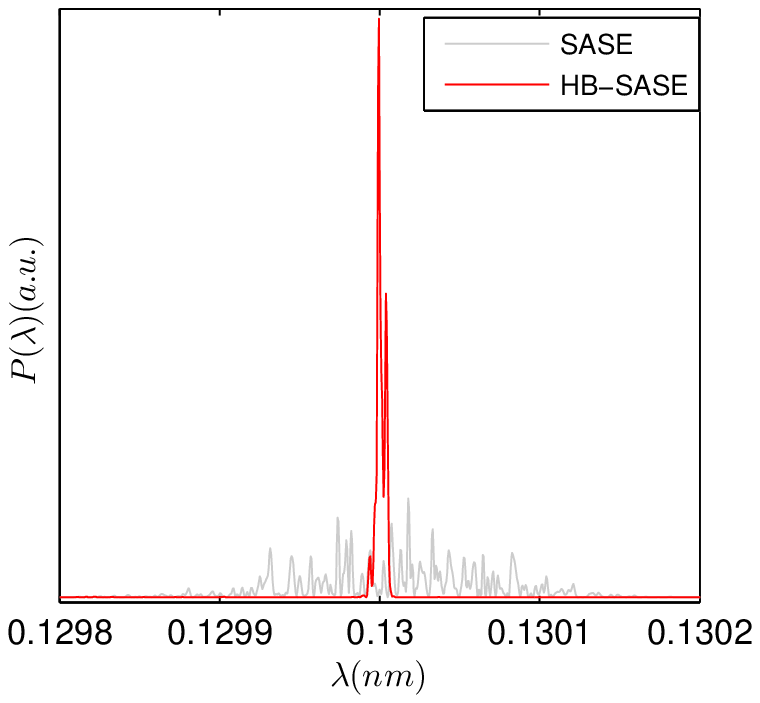}
\caption{Hard X-ray example at $\lambda_r=0.13$~nm: the pulse profiles and spectra of SASE and HB-SASE.}\label{LCLSPULSES}
\end{figure}

A practical example is now shown of HB-SASE applied to a hard X-ray FEL. The parameters are $\lambda_r$=0.13~nm, beam energy $E$=14.7~GeV, peak current  $I_{\mathrm{pk}}=3000$A, bunch charge $Q=10$~pC, $\lambda_u = 30$mm and $\rho=4.17\times 10^{-4}$. The delays are set so that at saturation the total slippage is the FWHM electron bunch length, giving $\bar{s}_1=1.08\bar{l}$. The undulator modules have length $\bar{l}=0.5$ equivalent to $L_u=2.85$m. The results are shown in S.I. units. Figure \ref{LCLSPULSES} shows the pulse profiles and spectra of SASE and HB-SASE at saturation. The efficacy of HB-SASE is clearly demonstrated---the SASE pulse is a chaotic sequence of phase uncorrelated spikes whereas the HB-SASE pulse is near single spike with slowly varying phase (not shown). For SASE the coherence time $t_{\mathrm{coh}}=l_{\mathrm{coh}}/c$ is 0.43fs, close to the expected $2\pi l_c/c=0.52$fs, and $\sigma_{\lambda}/\lambda=4.3\times 10^{-4}\simeq \rho$ as expected. For HB-SASE $t_{\mathrm{coh}}=7.15$~fs with $\sigma_{\lambda}/\lambda=2.0\times 10^{-5}$. The FWHM pulse durations and bandwidths give time-bandwidth product $\Delta \nu \Delta t = (1/\lambda)(\Delta \lambda/\lambda) c \Delta t = 32$ for SASE  and $\Delta \nu \Delta t=0.85$ for HB-SASE, indicating the HB-SASE output pulse is close to transform limited.

\begin{figure}
\includegraphics[width=42mm]{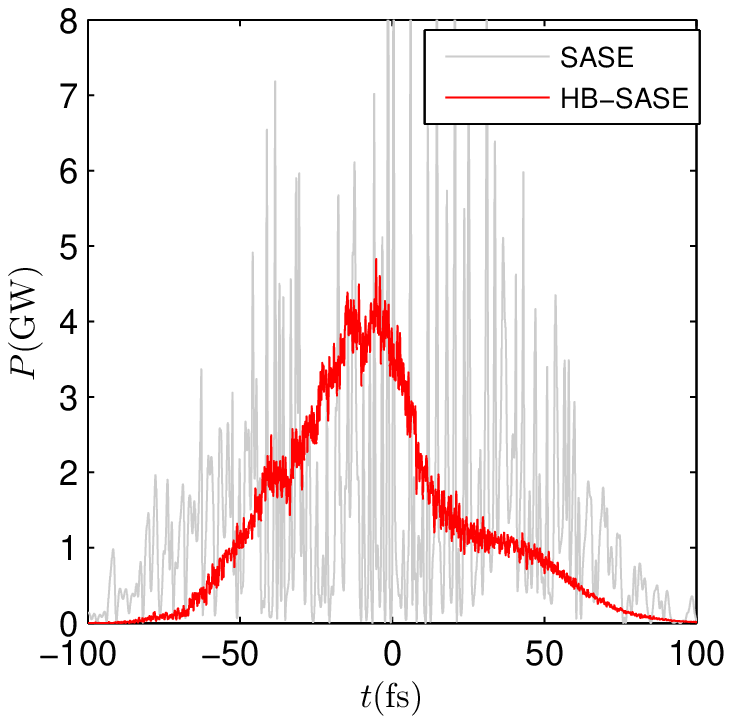}
\includegraphics[width=42mm]{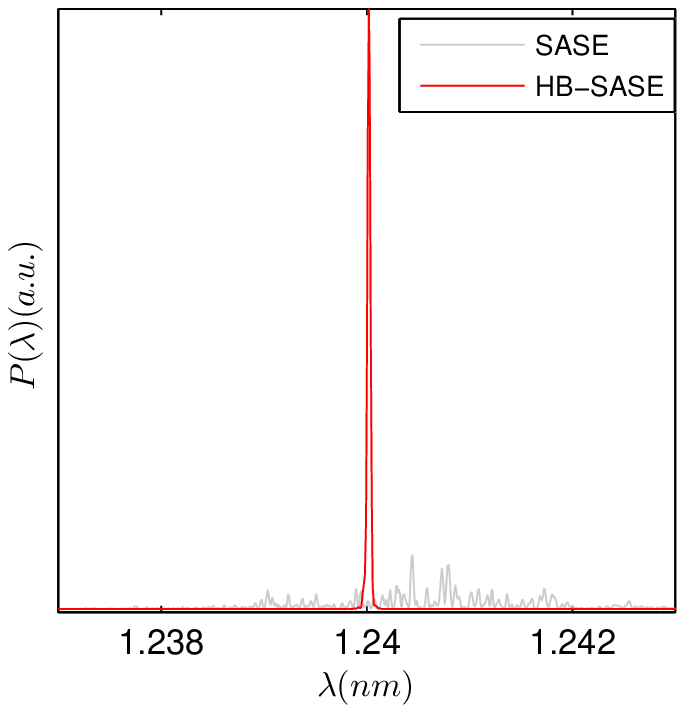}
\caption{Soft X-ray example at $\lambda_r=1.24$~nm.}\label{NLSPULSES}
\end{figure}

Finally, a brief example is shown of a soft X-ray FEL operating at $\lambda_r=1.24$~nm with $E$=2.25~GeV,  $I_{\mathrm{pk}}=1200$~A and $Q=200$~pC. Figure \ref{NLSPULSES} shows the results at saturation. For HB-SASE  $\Delta \nu \Delta t = 0.53$ indicating the output pulse is transform limited.

In this Letter isochronous delay-chicanes were assumed~\cite{CHICANES1} which do not disturb or enhance the rate of electron microbunching. Another option  would be to use standard non-isochronous chicanes (for example simple 4-dipole chicanes) which introduce positive momentum compaction upon the electrons,  plus occasional `correction' chicanes with a strong negative compaction \cite{CHICANES2} to cancel the accumulated positive  compaction. When non-isochronous chicanes only are used, the cooperation length is reduced~\cite{MODELOCKEDFEL} and for longer delays the momentum compaction shears the developing microbunching. This limits the coherence length increase to approximately one order of magnitude. Using isochronous chicanes, undulator section lengths of $\bar{l}\approx 0.5$ gave a HB-SASE coherence length more than two orders of magnitude greater than SASE.  Increasing  $\bar{l}$ further to 1.0 and 2.0 the maximum coherence lengths reduced to a factor approximately 50 and 10 times  respectively. As most current FELs have been designed so that module length $\bar{l}\gtrsim 1$, this suggests that HB-SASE may be successfully applied at existing facilities by inclusion of suitably compact chicanes between undulator sections.

The High-Brightness SASE FEL has been described which may generate transform-limited X-ray pulses. The method de-localises the collective FEL interaction to break the radiation coherence length dependence on the FEL cooperation length. While the technique described here was applied to a high-gain FEL, the similarity of the FEL process to other collective systems may open a route to new methods for coherence control and noise reduction in such systems.

\end{document}